\documentclass[a4paper,11pt,reqno]{amsart}

\usepackage{geometry}
\usepackage{graphicx}
\usepackage{indentfirst,csquotes}

\usepackage[shortlabels]{enumitem}
\usepackage{graphicx}
\usepackage{amsmath,amsthm,amssymb,mathtools,thmtools}
\numberwithin{equation}{section}
\usepackage{threeparttable}

\usepackage{physics2}
\usephysicsmodule{ab}
\usepackage{lscape}
\theoremstyle{plain}

\theoremstyle{definition}

\theoremstyle{remark}

\usepackage{CJKutf8} 

\usepackage{hyperref}
\usepackage[color,final]{showkeys}

\definecolor{refkey}{rgb}{0.1,0.1,0.7}
\definecolor{citekey}{rgb}{0.1,0.1,0.7}
\definecolor{labelkey}{rgb}{0.1,0.1,0.7}

\begin{document}
\renewcommand{\sc}{\scshape}

\title[Measurement precision in lenear dose-response relationships]{Statistical evaluation of measurement precision in linear dose-response relationships via interlaboratory studies}
\author[J. Takeshita]{Jun-ichi Takeshita${}^{1,2}$}
\author[Y. Ikeuchi]{Yuto Ikeuchi${}^3$}
\author[T. Suzuki]{Tomomichi Suzuki${}^2$}

\address{${}^1$Research Institute of Science for Safety and Sustainability, National Institute of Advanced Industrial Science and Technology (AIST), Tsukuba, Japan}
\email{jun-takeshita@aist.go.jp}
\address{${}^2$Department of Information Science and Technology,  Faculty of Information Science and Technology, Tokyo University of Science, Noda, Japan}
\address{${}^3$Department of Industrial and System Engineering, Graduate School of Science and Technology, Tokyo University of Science, Noda, Japan}

\date{\today}
\begin{abstract}
  This paper proposes a framework for evaluating the statistical precision of measurement methods from interlaboratory studies where the outcome is a dose–response relationship summarized by a regression line.
  For such measurement methods, where a linear mixed-effects model is applied that allows laboratories to differ in both baseline level and dose–response slope, we define precision evaluation metrics specified in ISO 5725, repeatability and between-laboratory variances.
  These are method-level precision metrics, and the latter are constructed as design-averaged dose-specific between-laboratory variances over the dose levels and the participating laboratories.
  For fully balanced designs with common dose levels and equal replication, we obtain an exact decomposition of the total sum of squares, closed-form analysis of variance (ANOVA) estimators of the precision variances, and three associated $F$-tests targeting (i) the overall dose–response trend, (ii) homogeneity of intercepts, and (iii) homogeneity of slopes across laboratories.
  This formulation enables precision to be quantified and estimated directly and supports an evaluation of whether between-laboratory discrepancies are caused primarily by baseline shifts or by differences in sensitivity, in contrast to fixed-effect comparisons that only detect the presence of differences.
  Furthermore, we analyze data obtained from an interlaboratory study on observations in bronchoalveolar lavage fluid from experiments involving the intratracheal administration of nanomaterials to rats, using the proposed method as a case study.
  
  \noindent
  \textbf{key words:} Interlaboratory precision, Linear mixed-effects models, Variance components, Dose–response regression
ISO 5725
\end{abstract}

\maketitle

\section{Introduction}
\label{sec:intro}

The assessment of measurement precision is a fundamental problem in both metrology and applied statistics, particularly when measurement methods are to be standardized and compared across laboratories. While classical precision assessment focuses on variability in single quantitative outcomes, many scientific and regulatory contexts use measurements that characterize dose-response relationships, where the response is inherently functional rather than scalar.
The ISO 5725 series~\cite{iso2023ISO572512025,iso2025ISO572522025} (see Appendix~\ref{sec:model-prec-meas-iso-} ) provides an internationally accepted framework for evaluating repeatability (within-laboratory) and between-laboratory precision through interlaboratory studies, assuming that each measurement yields a single quantitative value.
However, in fields such as medicine, toxicology, and environmental science, they often deal with a dose-response relationship estimated from multiple observations, for example in exposure–response modeling or toxicity studies~\cite{fda/cder2003GuidnaceIndustryExposure,kappenberg2023Guidancestatisticaldesign}.
The existing ISO framework cannot be applied straightforwardly to such data.

When a linear relationship is assumed, which is the simplest and easiest case, a common approach to comparing dose-response relationships across laboratories is the analysis of covariance (ANCOVA), in which laboratory effects are treated as fixed and differences among regression lines are assessed through hypothesis testing~\cite{laird2021Exposure­Response,shanks2026InterlaboratoryComparisonReal}.
While suitable for detecting systematic differences, this approach does not allow the estimation of between-laboratory variance components and therefore does not provide a precision assessment in the sense of ISO 5725. An alternative perspective was proposed by Uhlig et al.~\cite{uhlig2015ValidationQualitativePCR}, who translated the ISO concepts of repeatability and between-laboratory precision into the parameters of a probability-of-detection (POD) model for qualitative polymerase chain reaction (PCR) data.
In their framework, laboratory effects are treated as random; however, their model assumes the homogeneity of slopes across laboratories, motivated by empirical considerations in the specific PCR application, and also does not explicitly quantify repeatability precision.

In general interlaboratory settings, both the intercepts and slopes of dose-response relationships may vary across laboratories, reflecting differences in sensitivity, calibration, or experimental protocols.
Ignoring slope heterogeneity may lead to misleading conclusions about measurement precision, while treating laboratories as fixed effects makes it impossible to calculate variance-based precision metrics.
Thus, there is a need for a statistically coherent model that (i) accommodates random laboratory effects on both intercepts and slopes, (ii) allows the explicit definition and estimation of repeatability and between-laboratory variance components, and (iii) enables formal inference on the homogeneity of dose-response relationships.

This study proposes a model for assessing the precision of measurement methods that involve linear dose–response relationships.
Within this framework, we define repeatability and between-laboratory variances in a manner consistent with the conceptual foundations of ISO 5725, and develop estimation and hypothesis testing procedures for both regression coefficients and variance components.
The methodology is applied to analyze data from an actual interlaboratory study involving five laboratories, focusing on bronchoalveolar lavage fluid (BALF) examinations conducted during intratracheal administration studies of nanomaterials in rats.

\section{Proposed model}
\label{sec:prop-model}

\subsection{Definition of the model}
\label{sec:defin-model}

This study proposes the following mixed-effect model for evaluating measurement precision for dose-response data.
\begin{align}
  Y_{ij} = (a_{0} + A_{i}) + (b_{0} + B_{i}) x_{ij} + E_{ij}, \quad i=1,\ldots, m, \ j = 1,\ldots, n,
  \label{eq:1}
\end{align}
where $a_0$ and $b_0$ denote the true intercept and slope, respectively.
The terms $A_i$ and $B_i$ represent the laboratory-specific components of variation in the intercept and slope in laboratory $i$, with $A_i\sim N(0,\sigma_A^2)$ and $B_i\sim N(0,\sigma_B^2)$.
The residual error $E_{ij}$ satisfies $E_{ij}\sim N(0,\sigma_E^2)$.
All random variables $A_i$, $B_i$, and $E_{ij}$ are assumed to be independent.
Furthermore, we assume that the combinations of levels used in each laboratory are identical, that is, $x_{1j}=x_{2j}=\cdots=x_{mj}$.
Consequently, we omit the subscript $j$ from $x$ in what follows.
We also assume that $x_1,\ldots,x_m$ are centered, i.e. $(x_1+\cdots+x_m)/m=0$.

Hereafter, the following notation is used:
\begin{align}
  S_{xxL} &:=   \sum_{j=1}^m \left(x_j - \bar{x}\right)^2 =  \sum_{j=1}^m x^2_j, \quad 
  S_{xxT} := \sum_{i=1}^m \sum_{j=1}^n x_{j}^{2} = m \sum_{j=1}^n x_{j}^{2}, \label{eq:2} \\
  S_{xyL(i)}&:= \sum_{j=1}^n x_{j} Y_{ij}, \quad 
  S_{xyT} := \sum_{i=1}^m \sum_{j=1}^n x_{j} Y_{ij}, 
           \label{eq:3}
  \end{align}
Obviously, $S_{xxT} = m S_{xxL}$.

Next, let ${\hat{a}}_0$ and ${\hat{b}}_0$ denote the estimators of $a_0$ and $b_0$, respectively.
Specifically,
\begin{align}
  \hat{a}_{0} = \bar{\bar{Y}} = \frac{1}{mn} \sum_{i=1}^m \sum_{j=1}^n Y_{ij}\quad \text{and} \quad \hat{b}_{0} = \frac{S_{xyT}}{S_{xxT}}.
  \label{eq:4}
\end{align}
We define $\alpha_i$ as a statistic describing the difference between the intercept in laboratory $i$ and the overall intercept: 
\begin{align}
  \alpha_{i} = \bar{Y}_{i} - \hat{a}_{0} = \bar{Y}_{i} - \bar{\bar{Y}}, \quad \text{where } \bar{Y}_{i} = \frac{1}{n} \sum_{j=1}^n Y_{ij}.
  \label{eq:5} 
\end{align}
Similarly, we define $\beta_i$ as a statistic describing the difference between the slope in laboratory $i$ and the overall slope:
\begin{align}
   \beta_{i} = \frac{S_{xyL(i)}}{S_{xxL}} - \hat{b}_{0} = \frac{S_{xyL(i)}}{S_{xxL}} - \frac{S_{xyT}}{S_{xxT}}.
  \label{eq:6} 
\end{align}
Finally, let $\varepsilon_{ij}$ denote the residual under the proposed model, $\varepsilon_{ij}=Y_{ij}-{\hat{Y}}_{ij}$, where ${\hat{Y}}_{ij}=({\hat{a}}_0+\alpha_i)+({\hat{b}}_0+\beta_i)x_j$.

\subsection{Decomposition of the sum of squares}
\label{sec:decomp-sum-squar-mod}

For the conventional simple linear regression model, hereafter referred to as the simple model,
\begin{align}
  Y'_{ij} = a_{0} + b_{0}x_{ij} + E'_{ij}, \quad \text{where } E'_{ij} \sim N \left(0, \sigma^{2}_{E'}\right),
  \label{eq:7}
\end{align}
the decomposition of the total sum of squares is
\begin{align}
    \sum_{i=1}^{n} \sum_{j=1}^{m} \left(Y'_{ij} - \bar{\bar{Y}}\right)^2
    =\sum_{i=1}^{n} \sum_{j=1}^{m} \left(Y_{ij} - \hat{Y}'_{ij} \right)^2
    + \sum_{i=1}^{n} \sum_{j=1}^{m} \left(\hat{Y}'_{ij} - \bar{\bar{Y}} \right)^2,
  \label{eq:8}
\end{align}
where ${{\hat{Y}}^\prime}_i={\hat{a}}_0+{\hat{b}}_0x_{ij}$.
We summarize this decomposition as $S_{T'} = S_{E'} + S_{R}$, where $S_{T’}$ is the total sum of squares, $S_{E’}$ is the residual sum of squares, and $S_{R}$ is the regression sum of squares for the simple model.

Then, the total sum of squares around the overall mean can be decomposed as
  \begin{align}
    \sum_{i=1}^{n} \sum_{j=1}^{m} \left(Y_{ij} - \bar{\bar{Y}}\right)^2
    =\sum_{i=1}^{n} \sum_{j=1}^{m}\left(Y_{ij} - \hat{Y}_{ij} \right)^2
    + \sum_{i=1}^{n} \sum_{j=1}^{m} \left(\hat{Y}_{ij} - \hat{Y}'_{ij} \right)^2
    + \sum_{i=1}^{n} \sum_{j=1}^{m}\left(\hat{Y}'_{ij} - \bar{\bar{Y}} \right)^2.
    \label{eq:9}
  \end{align}
We summarize this decomposition as $S_{T} = S_{E} + S_{L} + S_{R}$.
The proof of the decomposition~\eqref{eq:9} is provided in Appendix~\ref{sec:proof-eq10}.

\subsection{ANOVA for the proposed model}
\label{sec:anova-for-prop-model}

\subsubsection{Definition of variance components}
\label{sec:defin-vari-comp}

  We define the repeatability variance ($\sigma_{r}^{2}$) and the between-laboratory variance ($\sigma_{L}^{2}$) for the proposed model as
  \begin{align}
    \sigma^{2}_{r} := \sigma^{2}_{E} \quad \text{and} \quad \sigma^{2}_{L} := \sigma^{2}_{A} + \frac{S_{xxL}}{n}\sigma^{2}_{B}. \label{eq:10}
  \end{align}

  We now comment on the definition of the between-laboratory variance.
  We define the laboratory effect $L_{ij}(x_j) := A_i+B_i x_j$.
  Then, we define a dose-specific between-laboratory variance
\begin{align}
  \tau^2_L(x_j) := \mathbb{V}[L_{ij}(x_j)] = \sigma^{2}_{A} + x_{j}^{2} \sigma^{2}_{B}.
  \label{eq:11}
\end{align}
Since the between-laboratory variance varies with the dose level, we define $\sigma_L^2$ as the mean of $\tau_L^2(x_j)$ over all $x_j$.
In other words,
\begin{align}
  \sigma^{2}_{L} := \frac{1}{mn} \sum_{i=1}^{n} \sum_{j=1}^{m} \mathbb{V}[L_{ij}(x_j)] = \sigma^{2}_{A} + \frac{S_{xxL}}{n}\sigma^{2}_{B}.
  \label{eq:12}
\end{align}
To explicitly distinguish the overall between-laboratory variance $\sigma_L^2$ from the dose-specific between-laboratory variance $\tau_L^2(x_j)$, we hereafter refer to the former as the design-averaged between-laboratory variance.

\subsubsection{Degrees of freedoms}
\label{sec:degr-freed}

Since, in the proposed model, the intercept and slope are estimated for each laboratory and the number of laboratories is $m$, a total of $2m$ parameters are estimated.
Consequently, the degrees of freedom for the error term are $mn-2m$.
The degrees of freedom for regression are $1$, and the total (overall) degrees of freedom are $mn-1$; therefore, the degrees of freedom for the between-laboratory are $2(m-1)(=(mn-1)-(mn-2m)-1)$.

\subsubsection{ANOVA tables for the proposed model}
\label{sec:anova-tabl-for-prop-}

An ANOVA table for the proposed model is shown in Table~\ref{tab:Basic-ANOVA-table}.
The expectations of the mean squares are provided in Appendix~\ref{sec:deriv-expect-sum-suq}.
From the ANOVA table, the repeatability and between-laboratory variances can be estimated as
\begin{align}
  \hat\sigma^2_r = V_E \quad \text{and} \quad \hat\sigma^2_L = \frac{2}{n}\left(V_L - V_E \right),
  \label{eq:13}
\end{align}
where $V_E=S_E/(mn-2m)$ and $V_L=S_L/(2(m-1))$.

Furthermore, $S_L$ can be expressed as
\begin{align}
  S_L &= \sum_{i=1}^m \sum_{j=1}^n \left(\hat{Y}_{ij} - \hat{Y}'_{ij} \right)^2
= \sum_{i=1}^m \sum_{j=1}^n \left(\alpha_i - \beta_i x_j \right)^2
  \label{eq:14} \\
      &= n \sum_{i=1}^m \alpha_i^2 + 2 \sum_{i=1}^m \alpha_i \beta_i \sum_{j=1}^n  x_j + \sum_{i=1}^m \beta^2_i \sum_{j=1}^n  x_j^2
        \label{eq:15}\\
  &= n \sum_{i=1}^m \alpha_i^2 + S_{xxL} \sum_{i=1}^m \beta^2_i.
  \label{eq:16}
\end{align}
In other words, the between-laboratory mean square can be decomposed into contributions from the slope and intercept that are described respectively by $S_A := n\sum_{i=1}^{n}\alpha_i^2$ and $S_B := S_{xxL}\sum_{i=1}^{n} \beta_i^2$. 
This is because the degrees of freedom associated with the slope and intercept are the same, that is, $m-1$.
These facts lead to a refined ANOVA table (Table~\ref{tab:Detailed-ANOVA-table} ), which we refer to as the detailed ANOVA table.
The expectations of the mean squares are also provided in Appendix~\ref{sec:deriv-expect-sum-suq}.
In contrast, we refer to the first ANOVA Table (Table~\ref{tab:Basic-ANOVA-table} ) as the basic ANOVA table.

\subsubsection{Three $F$-tests}
\label{sec:three-tests}

Using these ANOVA tables, we can perform three $F$-tests: (i) a significance test for the regression coefficients, (ii) a test of homogeneity of intercepts, and (iii) a test of homogeneity of slopes.

\paragraph{\emph{Significance test for the regression coefficients}}

We consider the null and alternative hypotheses $H_0: b_0 = 0, \ H_1: b_0 \neq 0$.
Since, from Table~\ref{tab:Detailed-ANOVA-table}, $\mathbb{E}[V_R] = b_0 S_{xxT} + \sigma_B^2 S_{xxL} + \sigma_E^2$ and $\mathbb{E}[V_B] = S_B / (m-1) = \sigma_B^2 S_{xxL} + \sigma_E^2$, it follows that $\mathbb{E}(V_R) = \mathbb{E}(V_{B})$ under $H_0$.
Therefore, the significance of the regression coefficients can be tested using the statistic  $V_R / V_B \sim F(1,m-1)$ under $H_0$.
We note that, in ordinary simple linear regression, the corresponding test statistic is $V_R / V_E$.

\paragraph{\emph{Test of homogeneity of intercepts}}

We consider the null and alternative hypotheses $H_0: \sigma^2_A = 0, \ H_1: \sigma^2_A \neq 0$.
Since, from Table~\ref{tab:Detailed-ANOVA-table}, $\mathbb{E}(V_A) = n \sigma^2_A + \sigma^2_E$, it follows that $\mathbb{E}(V_A) = \mathbb{E}(V_E)$ under $H_0$.
Therefore, the homogeneity of intercepts can be tested using the statistic  $V_A / V_E \sim F(m-1,mn-2m)$ under $H_0$.

\paragraph{\emph{Test of homogeneity of slopes}}

We consider the null and alternative hypotheses $H_0: \sigma^2_B = 0, \ H_1: \sigma^2_B \neq 0$.
Since, from Table~\ref{tab:Detailed-ANOVA-table}, $\mathbb{E}(V_B) = \sigma^2_B S_{xxL}+ \sigma^2_E$, it follows that $\mathbb{E}(V_B) = \mathbb{E}(V_E)$ under $H_0$.
Therefore, the homogeneity of intercepts can be tested using the statistic  $V_B / V_E \sim F(m-1,mn-2m)$ under $H_0$.




\section{Real example} \label{sec:real-example-1}

\subsection{Data description} \label{sec:data-descr}

This study uses data from an interlaboratory study, involving five participating laboratories, concerning BALF examinations conducted during intratracheal administration studies of nanomaterials in rats~\cite{aist2018AnnualReportProject}.
Intratracheal administration is an animal testing method in which a test sample is administered directly into the trachea of a rat~\cite{driscoll2000IntratrachealInstillationas}.
A BALF test involves using a bronchoscope to inject sterile saline into a portion of the lung, recovering the fluid, and analyzing it to assess lung condition~\cite{davidson2020Bronchoalveolarlavageas}.

Each laboratory examined four dose levels—--$1,\text{mg/kg}, 0.33,\text{mg/kg}, 0.10,\text{mg/kg}$, and a control---with five rats assigned to each dose group. Because the doses were set in a geometric series with a common ratio of $1/\sqrt{10}$ from the highest dose, we treat these four doses as $1$,$1/\sqrt{10}$,$0.1$, and $0.1/\sqrt{10}$ in the analysis, and we use the common logarithm of these dose values.
Moreover, to apply the proposed method, the dose variable must be centered. 
After centering, the log-transformed doses become $-0.75, -0.25, 0.25$, and $0.75$.

Because the rat strain and age were standardized across all laboratories, we assume that individual differences were minimized as much as possible; thus, each dose level effectively has five replicates.
The target organ for observation was the right lung, and the parameters measured from the collected fluid were total cell count, neutrophil count, neutrophil proportion, lactate dehydrogenase (LDH), and total protein.
Of these, we analyze the natural log-transformed LDH and total protein, as these variables are quantitative.

\subsection{Result of the example}
\label{sec:result-example}

Table~\ref{tab:res-data} shows the estimated intercepts and slopes for the five laboratories.
Tables~\ref{tab:bANOVA-LDH} and~\ref{tab:dANOVA-LDH} present the basic and detailed ANOVA tables, respectively, for the LDH. 
From Table~\ref{tab:bANOVA-LDH}, the estimated repeatability and between-laboratory variances are $0.119$ and $1.35$, respectively.
Table~\ref{tab:dANOVA-LDH} indicates that both the intercept and the slope are statistically significant; in particular, the intercept is highly significant, with a very large $F$-value.

Tables~\ref{tab:bANOVA-TP} and~\ref{tab:dANOVA-TP} present the basic and detailed ANOVA tables, respectively, for the total protein.
From Table~\ref{tab:bANOVA-TP}, the estimated repeatability and between-laboratory variances are 0.086 and 0.025, respectively.
Table~\ref{tab:dANOVA-TP} indicates that only the intercept is statistically significant.

\section{Discussion and concluding remarks}
\label{sec:disc-concl-remarks}

In this study, we extend the ISO 5725 precision assessment framework from single measured values to dose–response relationships.
Whereas ISO 5725 assumes that each result is a single scalar measurement, our focus is on measurement results summarized by a regression line.
We propose a linear mixed model with explicitly defined variance components for repeatability (within-laboratory precision) and between-laboratory precision.
Based on a decomposition of the sum of squares, we derive ANOVA-style estimators and three associated $F$-tests for (i) the significance of the regression coefficients, (ii) the homogeneity of intercepts, and (iii) the homogeneity of slopes.
This framework enables the direct evaluation of precision as variance components, which is not possible under conventional ANCOVA that treats a laboratory as a fixed effect.

In the ISO 5725 setting, repeatability and between-laboratory variances are defined under the assumption that the measurement results are single quantities.
To extend this to regression-based results, we decompose between-laboratory variability into two components: an intercept component and a slope component.
To avoid ambiguity, we use the term ``between-laboratory variance'' for the design-averaged quantity $\sigma_L^2$ and use ``dose-specific between-laboratory variance'' for the variance at a given dose level $\tau_L^2(x)$.
Their definitions are reposted below.

Because the dose-specific between-laboratory variance depends on dose level, it must be aggregated to obtain a single precision metric.
We define the dose-specific between-laboratory variance as $\tau_L^2(x) := Var(A_i + B_i x) = \sigma_A^2 + \sigma_B^2 x^2$.
We then adopt the design-averaged between-laboratory variance as the average of $\tau_L^2(x)$ over the set of dose levels used in the study design: $\sigma_L^2 := (1/(mn)) \sum_{i=1}^m \sum_{j=1}^n \tau_L^2(x_j) = \sigma_A^2 + (S_{xxL} / n) \sigma_B^2$.
Note that, since this study assumes that the setting of dose levels is identical across all participating laboratories, the definition of $\sigma_L^2$ is also the average across all laboratories: $\sigma_L^2 := (1/(mn))\sum_{i=1}^m \sum_{j=1}^n \tau_L^2(x_j)$.
This design-averaged between-laboratory variance summarizes the overall between-laboratory variation of the measurement method for the prescribed design and is well suited to standardization and specification development, where the method is evaluated as a single unified procedure under defined conditions. 
Moreover, from $\tau_L^2(x)=\sigma_A^2+\sigma_B^2 x^2$, it is evident that the dose-specific between-laboratory variance increases with $|x|$.
Consequently, for the same measurement method with the same $\sigma_B^2$, designs that use a wider spread of dose levels (larger average $x^2$) will yield a larger design-averaged between-laboratory variance $\sigma_L^2$.

If the objective is to detect between-laboratory differences in the dose–response relationship, ANCOVA can be used to test for differences in intercepts and slopes as fixed effects.
However, because the laboratory is treated as a fixed factor, ANCOVA cannot estimate between-laboratory variance and therefore cannot address measurement precision as a variance-based metric in the sense of ISO 5725.
ANCOVA can indicate whether differences exist, but it cannot quantify the magnitude of inherent between-laboratory variability.
By contrast, our framework models both intercepts and slopes as random effects and, through ANOVA-based tests, allows the separate evaluation of heterogeneity in intercepts and slopes.

A key advantage of our approach is that, under a fully balanced interlaboratory design with common dose levels and equal replication, the proposed variance-component estimators and associated tests admit transparent, closed-form ANOVA expressions derived from the sum-of-squares decomposition.
This yields easily interpretable precision metrics aligned with the ISO 5725 framework and facilitates communication in standardization contexts.
At the same time, the proposed model is naturally viewed as a linear mixed-effects model; thus, when the design is unbalanced due to missing observations, unequal replication, or partially mismatched dose levels, the analysis can be extended using likelihood-based estimation, such as the restricted maximum likelihood (REML) approach.
In this sense, the present framework provides a coherent bridge between the classical ANOVA tradition (balanced designs with closed-form inference) and modern mixed-model practice (general unbalanced designs with likelihood-based inference).

The basic decomposition $S_T=S_R+S_L+S_E$ clearly partitions variability into three components: (i) within-laboratory/residual error, (ii) between-laboratory differences in intercepts and slopes, and (iii) the overall dose–response relationship.
A more detailed ANOVA further decomposes the between-laboratory component as$ S_L=S_A+S_B$, where $S_A$ captures between-laboratory differences in intercepts and $S_B$ captures those in slopes.
This decomposition quantifies whether discrepancies across laboratories arise primarily from shifts in reference values (intercept differences) or from differences in sensitivity (slope differences).

The use of the ratio $V_R/V_B$ in the significance test for the regression coefficients, which differs from standard simple regression, reflects the fact that, when slopes vary across laboratories, the explanatory power of the regression is affected by both the residual error $\sigma_E^2$ and the random variation in slopes $\sigma_B^2$.
Thus, the proposed test evaluates whether the overall dose–response relationship is statistically significant while explicitly accounting for between-laboratory heterogeneity in slope.

Analysis of the actual data showed that, for LDH, both the intercept and the slope were significant, with the intercept making a particularly large contribution.
This suggests that non–dose-dependent factors—such as differences in the overall baseline (background levels) or systematic deviations in the measurement process (sample handling, instrument calibration, or trial runs) —–may be major drivers of between-laboratory variation.
In contrast, for total protein, only the intercept was significant and the slope was homogeneous.
This pattern indicates that in the measurement method, the sensitivity (dose–response) is relatively stable across laboratories, but baseline shifts remain.
In summary, the proposed method does more than simply conclude that ``between-laboratory variation exists.''
It decomposes that variation into intercept and slope components and thereby reveals its underlying structure.
This, in turn, enables direct and concrete links to improvement strategies for the measurement method, such as calibration adjustments, protocol standardization

While the framework proposed in this study is conceptually simple and easy to interpret, it relies on several assumptions.

(i) \emph{Assumption of a linear dose–response relationship.}
This paper focuses on linear dose–response relationships, even though nonlinear models (such as the maximum effect (Emax)  or logistic models) are often more appropriate in toxicity and pharmacodynamic studies~\cite{parodi1989ResultsAnimalStudies,proost2020PopulationPharmacodynamicModeling}.
A linear model is useful as a first approximation, but substantial nonlinearity may lead to a lack of fit in certain dose regions.
Extending the present framework to nonlinear dose–response models is therefore an important direction for future work.

(ii) \emph{Assumption of common dose levels and equal replication (fully balanced design)}.
We assume that all laboratories use the same set of dose levels and that the number of replicates per dose is the same.
These assumptions allow a simple ANOVA decomposition and straightforward calculation of degrees of freedom.
In practice, however, missing data and imbalance are common.
In such situations, it is necessary to extend the analysis to REML or other likelihood-based methods and to use approximate degrees of freedom (e.g., Satterthwaite-type approximations) that properly account for imbalance.

(iii) \emph{Assumptions of normality and independence}.
We assume normality and mutual independence of the random effects and error terms.
In particular, we assume that $A_i$ and $B_i$ are independent for simplicity.
When differences in calibration conditions or experimental procedures affect both the intercept and the slope, a model that allows correlation between $A_i$ and $B_i$ may be more realistic.
Developing and evaluating such correlated random-effects extensions is a relevant topic for future research.

The framework proposed in this study can be extended in several natural directions:

(i) \emph{Models with correlated random effects ($A_i$ and $B_i$)}.
Allowing covariance between the random intercept ($A_i$) and slope ($B_i$) would enable the model to capture trade-offs, such as laboratories with higher sensitivity also having higher baselines.

(ii) \emph{Handling unbalanced designs and missing data}.
Because missing observations and unequal replicate counts are common in interlaboratory studies, it is important to link the proposed framework to likelihood-based estimation and hypothesis testing (e.g., REML) that can accommodate imbalance while preserving valid inference.

(iii) \emph{Generalization to nonlinear dose–response relationships}.
In many toxicology and pharmacology applications, nonlinear mixed-effects models (e.g., Emax, logistic) are more appropriate than linear models.
Extending the precision concepts introduced here (repeatability and between-laboratory variance) to nonlinear dose–response models is a key avenue for future research.

(iv) D\emph{ose-region–specific precision metrics}.
In this paper, we summarize between-laboratory precision using the design-averaged between-laboratory variance $\sigma_L^2$.
In regulatory settings, however, precision within specific dose regions (e.g., near the no observed adverse effect level) may be of primary interest.
Dose-weighted averaging that emphasizes critical regions, along with precision profiles $\tau_L^2(x)$ plotted across dose levels, would provide more tailored, practically relevant precision metrics.

This study proposes variance component definitions, estimation methods, and testing procedures consistent with the ISO 5725 concept of precision, specifically for interlaboratory studies involving dose–response data.
Unlike approaches based solely on fixed effects, the proposed framework evaluates the precision of a measurement method directly in terms of variance and decomposes between-laboratory differences into intercept and slope components, enabling clear interpretation of their respective contributions.
The case studies demonstrate that this framework is practical and can yield meaningful insights into the structure of between-laboratory variation in real data.

\section{Tables}
\label{sec:tabl}

\begin{table}[htbp]\centering
  \small
\caption{Basic ANOVA table}
\label{tab:Basic-ANOVA-table}
{\everymath{\displaystyle}
\begin{threeparttable}
\begin{tabular}{ccccc}
\hline
  factor & $S$ & $\phi$ & $V$  & $\mathbb{E}[V]$\\\hline
  Between-Laboratory: $L$ & $S_{L}$ & $\phi_{L} = 2(m-1)$
                        & $V_{L} = S_L / \phi_L$  & $(n / 2) \sigma^2_L + \sigma^2_E$\\
  Regression: $R$ & $S_R$ & $\phi_R = 1$ & $V_R = S_R$
                              &  $B^2_0 S_{xxT} + \sigma^2_B S_{xxL} + \sigma^2_E$\\
  Radisual: $E$ & $S_E$ & $\phi_E = mn - 2m$ &
                                               $V_E = S_E / \phi_E$ & $\sigma_E$\\\hline
  Total: $T$ & $S_T$ & $\phi_T = mn-1$ &&\\\hline
\end{tabular}
\begin{tablenotes}
  \item   $S_L = \sum_{i=1}^{n} \sum_{j=1}^{m} \left(\hat{Y}_{ij} - \hat{Y}'_{ij} \right)^2$, $S_R = \sum_{i=1}^{n} \sum_{j=1}^{m}\left(\hat{Y}'_{ij} - \bar{\bar{Y}} \right)^2$,
    \item $S_E = \sum_{i=1}^{n} \sum_{j=1}^{m} \left(Y_{ij} - \bar{\bar{Y}}\right)^2$, $S_T = \sum_{i=1}^{n} \sum_{j=1}^{m}\left(Y_{ij} - \hat{Y}_{ij} \right)^2$
\end{tablenotes}
\end{threeparttable}
}
\end{table}

\begin{table}[htbp]\centering
  \small
\caption{Detailed ANOVA table}
\label{tab:Detailed-ANOVA-table}
{\everymath{\displaystyle}
\begin{threeparttable}
  \begin{tabular}{ccccc}
\hline
  factor & $S$ & $\phi$ & $V$ & $\mathbb{E}[V]$\\\hline
  Intercept: $A$ & $S_A$ & $\phi_A = m-1$ & $V_A = S_A / \phi_A$ & $ n \sigma^2_A + \sigma^2_E$\\
 Slope: $B$ & $S_B$ &  $\phi_B = m-1$ & $V_B = S_B / \phi_B$ & $ \sigma^2_B S_{xxL} + \sigma^2_E$\\
  Regression: $R$ & $S_R$ & $\phi_R = 1$ & $V_R = S_R$ & $b^2_0 S_{xxT} + \sigma^2_B S_{xxL} + \sigma^2_E$\\
  Radisual: $E$ & $S_E$ & $\phi_E = mn - 2m$ & $V_E = S_E / \phi_E$ & $\sigma_E$\\\hline
  Total: $T$ & $S_T$ & $\phi_T = mn-1$ &&\\\hline
\end{tabular}
\begin{tablenotes}
  \item $S_A = n \sum_{i=1}^{n} \alpha_i^2$, $S_B := S_{xxL} \sum_{i=1}^{n} \beta_i^2$, 
  \item $S_R = \sum_{i=1}^{n} \sum_{j=1}^{m}\left(\hat{Y}'_{ij} - \bar{\bar{Y}} \right)^2$, $S_E = \sum_{i=1}^{n} \sum_{j=1}^{m} \left(Y_{ij} - \bar{\bar{Y}}\right)^2$,
  \item $S_T = \sum_{i=1}^{n} \sum_{j=1}^{m}\left(Y_{ij} - \hat{Y}_{ij} \right)^2$
\end{tablenotes}
\end{threeparttable}
}
\end{table}

\begin{table}[htbp]
  \centering
  \small
\caption{LDH data (U/L)}
\label{tab:LDH-data}
\begin{tabular}{rrrrrr}
\hline
dose & Lab A & Lab B & Lab C & Lab D & Lab E \\
\hline
$0$   & $49$  & $46$  & $691$ & $39$  & $5$   \\
$0$   & $66$  & $33$  & $547$ & $48$  & $23$  \\
$0$   & $41$  & $28$  & $486$ & $43$  & $12$  \\
$0$   & $36$  & $36$  & $515$ & $40$  & $9$   \\
$0$   & $37$  & $37$  & $422$ & $44$  & $25$  \\
$0.1$ & $77$  & $123$ & $750$ & $67$  & $26$  \\
$0.1$ & $109$ & $83$  & $756$ & $54$  & $20$  \\
$0.1$ & $102$ & $98$  & $741$ & $73$  & $49$  \\
$0.1$ & $58$  & $93$  & $654$ & $61$  & $30$  \\
$0.1$ & $80$  & $199$ & $832$ & $75$  & $37$  \\
$0.33$& $159$ & $207$ & $1395$& $85$  & $72$  \\
$0.33$& $196$ & $186$ & $1418$& $120$ & $67$  \\
$0.33$& $192$ & $218$ & $1310$& $114$ & $87$  \\
$0.33$& $159$ & $209$ & $1458$& $118$ & $82$  \\
$0.33$& $148$ & $178$ & $1717$& $108$ & $90$  \\
$1$   & $147$ & $288$ & $1748$& $141$ & $205$ \\
$1$   & $196$ & $137$ & $1740$& $122$ & $47$  \\
$1$   & $182$ & $176$ & $1787$& $139$ & $55$  \\
$1$   & $232$ & $198$ & $1686$& $197$ & $129$ \\
$1$   & $319$ & $222$ & $1375$& $159$ & $148$ \\
\hline
\end{tabular}

\end{table}
\begin{table}[htbp]
\centering
  \small
\caption{Total protein data (mg/dL)}
\label{tab:TP-data}
\begin{tabular}{rrrrrr}
\hline
dose & Lab A49 & Lab B & Lab C & Lab D & Lab E \\
\hline
$0$    & $68$  & $7.4$  & $153.6$ & $39$  & $13.7$ \\
$0$    & $106$ & $7.0$  & $82.5$  & $48$  & $18.4$ \\
$0$    & $64$  & $7.7$  & $69.3$  & $43$  & $15.4$ \\
$0$    & $63$  & $8.7$  & $64.5$  & $40$  & $7.1$  \\
$0$    & $75$  & $8.2$  & $48.5$  & $44$  & $10.8$ \\
$0.1$  & $148$ & $21$   & $114.7$ & $67$  & $22.3$ \\
$0.1$  & $201$ & $18.9$ & $134.5$ & $54$  & $18.4$ \\
$0.1$  & $177$ & $18.3$ & $147.4$ & $73$  & $17.9$ \\
$0.1$  & $102$ & $19.2$ & $117.1$ & $61$  & $25.2$ \\
$0.1$  & $151$ & $29.3$ & $174.1$ & $75$  & $8.1$  \\
$0.33$ & $301$ & $33$   & $188.6$ & $85$  & $54.3$ \\
$0.33$ & $372$ & $47.1$ & $231.6$ & $120$ & $38.4$ \\
$0.33$ & $380$ & $32.7$ & $178.3$ & $114$ & $37.4$ \\
$0.33$ & $327$ & $32.8$ & $249.2$ & $118$ & $46.3$ \\
$0.33$ & $264$ & $31.9$ & $267.2$ & $108$ & $31.6$ \\
$1$    & $273$ & $54$   & $332.9$ & $141$ & $55.3$ \\
$1$    & $462$ & $20.7$ & $407.6$ & $122$ & $37.6$ \\
$1$    & $407$ & $28.1$ & $242.7$ & $139$ & $62.6$ \\
$1$    & $407$ & $39.2$ & $319.2$ & $197$ & $48.4$ \\
$1$    & $591$ & $41.6$ & $246.4$ & $159$ & $41.9$ \\
\hline
\end{tabular}
\end{table}

\begin{table}[htbp]
\centering
  \small
\caption{Estimates of intercepts and slopes for five laboratories}
\label{tab:res-data}
\begin{tabular}{crrrrrr}
\hline
&  & Lab A & Lab B & Lab C & Lab D & Lab E \\
\hline
LDH &  Intercept & $4.67$ & $4.72$ & $6.89$ & $4.88$ & $3.73$\\
& Slope & $1.07$ & $1.14$ & $0.82$ & $1.29$ & $1.43$\\\hline
Total protein &  Intercept & $4.67$ & $4.72$ & $6.89$ & $4.88$ & $3.73$\\
& Slope & $1.07$ & $1.14$ & $0.82$ & $1.29$ & $1.43$\\  
 \hline
\end{tabular}
\end{table}

\begin{table}[htbp]
  \centering
  \caption{Basic ANOVA table for LDH data}
  \label{tab:bANOVA-LDH}
\begin{tabular}{cccc}
\hline
Factor & $S$ & $\phi$ & $V$ \\
\hline
Between-labratory: $L$ & $S_L = 109$ & $\phi_L = 8$ & $V_L = 13.6$ \\
Regression: $R$ & $S_R = 41.51$ & $\phi_R = 1$ & $V_R = 41.51$ \\
Radisual: $E$ & $S_E = 10.69$ & $\phi_E = 90$ & $V_E = 0.12$\\\hline
$T$ & $S_T = 161.18$ & $\phi_T = 99$ &  \\
\hline
\end{tabular}

\end{table}
\begin{table}[htbp]
  \centering
  \caption{Detailed ANOVA table for LDH data}
  \label{tab:dANOVA-LDH}
  \begin{threeparttable}
\begin{tabular}{ccccc}
\hline
Factor & $S$ & $\phi$ & $V$ & $F_0$ \\
\hline
Intercept: $A$ & $S_A = 107.64$ & $\phi_A = 4$ & $V_A = 26.91$ & $227^{*}$ \\
Slope: $B$ & $S_B = 1.34$ & $\phi_B = 4$ & $V_B = 0.34$ & $2.82^{*}$ \\
Regression: $R$ & $S_R = 41.51$ & $\phi_R = 1$ & $V_R = 41.51$ & $124^{*}$ \\
Radisual: $E$ & $S_E = 10.69$ & $\phi_E = 90$ & $V_E = 0.12$ &  \\\hline
$T$ & $S_T = 161.18$ & $\phi_T = 99$ &  &  \\
\hline
\end{tabular}
  \begin{tablenotes}
  \item ${}^{*}$ indicates significance at the 5\%
\end{tablenotes}
\end{threeparttable}
\end{table}

\begin{table}[htbp]
  \centering
  \caption{Basic ANOVA table for total protein data}
  \label{tab:bANOVA-TP}
  \begin{tabular}{cccc}
\hline
Factor & $S$ & $\phi$ & $V$ \\
\hline
Between-labratory: $L$ & $S_L = 2.67$ & $\phi_L = 8$ & $V_L = 0.33$ \\
Regression: $R$ & $S_R = 36.26$ & $\phi_R = 1$ & $V_R = 36.26$\\
Radisual: $E$ & $S_E = 7.76$ & $\phi_E = 90$ & $V_E = 0.086$   \\\hline
Total: $T$ & $S_T = 46.69$ & $\phi_T = 99$ &    \\
\hline
\end{tabular}
\end{table}

\begin{table}[htbp]
  \centering
  \caption{Detailed ANOVA table for total protein data}
  \label{tab:dANOVA-TP}
\begin{threeparttable}
  \begin{tabular}{ccccc}
\hline
Factor & $S$ & $\phi$ & $V$ & $F_0$ \\
\hline
Intercept: $A$ & $S_A = 2.07$ & $\phi_A = 4$ & $V_A = 0.52$ & $5.996^{*}$ \\
Slope: $B$ & $S_B = 0.60$ & $\phi_B = 4$ & $V_B = 0.16$ & $1.739$ \\
Regression: $R$ & $S_R = 36.26$ & $\phi_R = 1$ & $V_R = 36.26$ & $242^{*}$\\
Radisual: $E$ & $S_E = 7.76$ & $\phi_E = 90$ & $V_E = 0.086$  & \\\hline
Total: $T$ & $S_T = 46.69$ & $\phi_T = 99$ &  &  \\
\hline
  \end{tabular}
  \begin{tablenotes}
  \item ${}^{*}$ indicates significance at the 5\%
\end{tablenotes}
\end{threeparttable}
\end{table}

\clearpage

\bibliographystyle{siam}
\bibliography{PrecisionForDR}

\section{Acknowledgements}
\label{sec:ackn}

When preparing this manuscript, the authors utilized Grammarly (version 1.164.0) for English grammar support and Microsoft 365 Copilot LaTeX formatting assistance.
The authors thoroughly reviewed and edited the manuscript and take full responsibility for the content.

\section{Conflict of Interest}
\label{sec:confl-inter}

The authors declare that there are no conflicts of interest regarding the publication of this article.

\clearpage

\appendix

\section{Proof of Eq.~\eqref{eq:9}}
\label{sec:proof-eq10}

Because the decomposition of the total sum of squares for the simple model~\eqref{eq:8} is well known, we use it without a proof.
From its application, it follows that
\begin{align}
  S_T &=  \sum_{i=1}^m \sum_{j=1}^n \left(Y_{ij} - \hat{Y}'_{ij} \right)^2 + S_R
        \label{eq:17}\\
      &=\sum_{i=1}^m \sum_{j=1}^n \left(Y_{ij} - \hat{Y}_{ij} + \hat{Y}_{ij} - \hat{Y}'_{ij} \right)^2 + S_R
        \label{eq:18}\\
      &=S_A + S_B + S_R
  + 2 \sum_{i=1}^m \sum_{j=1}^n \left(Y_{ij} - \hat{Y}_{ij}\right)\left(\hat{Y}_{ij} - \hat{Y}'_{ij} \right).
    \label{eq:19} 
\end{align}
Hence, we should show $\sum_{i=1}^m \sum_{j=1}^n (Y_{ij} - \hat{Y}_{ij})(\hat{Y}_{ij} - \hat{Y}'_{ij}) = 0$.
Since $\hat{Y}_{ij} - \hat{Y}'_{ij} = \alpha_i + \beta_i x_j$, we have 
\begin{align}
  \sum_{i=1}^m \sum_{j=1}^n \left(Y_{ij} - \hat{Y}_{ij}\right)\left(\hat{Y}_{ij} - \hat{Y}'_{ij} \right)
  = \sum_{i=1}^m \sum_{j=1}^n \alpha_i \left(Y_{ij} - \hat{Y}_{ij}\right) + \sum_{i=1}^m \beta_i \sum_{j=1}^n x_j  (Y_{ij} - \hat{Y}_{ij}).
    \label{eq:20}
\end{align}
Here, it is easy to show $\sum_{j=1}^n \alpha_i = 0$ and $\sum_{i=1}^m \beta_i = 0$.
Thus, we obtain the decomposition $S_T = S_A + S_B + S_R$.

\section{Derivation of the expectations of the sum of squares}
\label{sec:deriv-expect-sum-suq}

\subsection{Derivation of $\mathbb{E}(V_A)$}
\label{sec:V_A}

Since $\alpha_i = \bar{Y}_i - \bar{\bar{Y}}$, it holds that
\begin{align}
  \alpha_i
  = \left(a_0 + A_i + \bar{E}_i\right) - \left(a_0 + \bar{A} + \bar{\bar{E}}\right)
  =  \left(A_i -\bar{A}\right) + \left(\bar{E}_i -\bar{\bar{E}}\right),
    \label{eq:21}
\end{align}
where $\bar{A} = \sum_{i=1}^m A_i / m$, $\bar{E}_i = \sum_{j=1}^n E_{ij} / n$, and $\bar{\bar{E}} = \sum_{i=1}^m \bar{E}_i / m$.
From $A_i$ and $E_{ij}$ are independent,
\begin{align}
  \mathbb{E}\left[ S_A \right] &=  \mathbb{E}\left[n \sum_{i=1}^m \alpha_i^2\right]
                              \label{eq:22}\\
                            &= n \mathbb{E}\left[\sum_{i=1}^m \left(A_i -\bar{A}\right)^2 + \left(\bar{E}_i -\bar{\bar{E}}\right)^2\right]
                              \label{eq:23}
                              \\
   &= n(m-1)\sigma^2_A + (m-1)\sigma^2_E,
     \label{eq:24}
\end{align}
thus
\begin{align}
  \mathbb{E}\left[V_A\right] =  \frac{\mathbb{E}\left[S_A\right]}{\phi_A} = n\sigma^2_A + (m-1) \sigma^2_E.
                             \label{eq:25}
\end{align}

\subsection{Derivation of $\mathbb{E}(V_B)$}
\label{sec:V_B}

Since $\beta_i = S_{xyL(i)}/S_{xxL} - S_{xyT}/S_{xxT}$, it holds that
\begin{align}
  b_i &= \frac{1}{S_{xxL}}\sum_{j=1}^n x_j Y_{ij} - \frac{1}{S_{xxT}} \sum_{i=1}^m \sum_{j=1}^n x_j Y_{ij}
        \label{eq:26}\\
      &= \frac{1}{S_{xxL}} \left\{ \sum_{j=1}^n x_j \left[ a_0 + A_i (b_0 + B_i) x_j E_{ij}\right]\right\}
        - \frac{1}{S_{xxT}} \left\{ \sum_{i=1}^m \sum_{j=1}^n x_j \left[ a_0 + A_i (b_0 + B_i) x_j E_{ij}\right]\right\}
        \label{eq:27}\\
  &= \frac{1}{S_{xxL}} \left[ \sum_{j=1}^n (b_0 + B_i) x_j^2 + \sum_{j=1}^n x_j E_{ij}\right]
        - \frac{1}{S_{xxT}} \left[ \sum_{i=1}^m\sum_{j=1}^n (b_0 + B_i) x_j^2 + \sum_{i=1}^m\sum_{j=1}^n x_j E_{ij}\right]
    \label{eq:28}\\
  &= \frac{1}{S_{xxL}} \left[ (b_0 + B_i) S_{xxL} + \sum_{j=1}^n x_j E_{ij}\right]
        - \frac{1}{S_{xxT}} \left[ b_0 S_{xxT} + \bar{B} S_{xxT} + \sum_{i=1}^m\sum_{j=1}^n x_j E_{ij}\right]
    \label{eq:29}\\
    &= \left(B_i - \bar{B}\right)  + \left(\frac{\sum_{j=1}^n x_j E_{ij}}{S_{xxL}} - \frac{\sum_{i=1}^m\sum_{j=1}^n x_j E_{ij}}{S_{xxT}} \right),
      \label{eq:30}
\end{align}
where $\bar{B} \sum_{i=1}^m B_i / m$.
Here, let $U_i:= \sum_{j=1}^n x_j E_{ij} / S_{xxL}$, it follows that $\sum_{i=1}^m\sum_{j=1}^n x_j E_{ij} / S_{xxT} = \sum_{i=1}^m U_i / m$, which is denoted by $\bar{U}$;
therefore, \eqref{eq:30} becomes
\begin{align}
  b_i = \left(B_i - \bar{B}\right)  + \left(U_i - \bar{U} \right).
  \label{eq:31}
\end{align}

Since 
\begin{align}
  \mathbb{V} [U_i] = \frac{1}{S_{xxL}^2} \mathbb{V} \left[\sum_{i=1}^m x_jE_{ij}\right] =  \frac{1}{S_{xxL}^2} \sum_{i=1}^m  x_i^2 \mathbb{V} [E_{ij}] = \frac{\sigma^2_E}{S_{xxL}},
  \label{eq:32} 
\end{align}
from Eqs.~\eqref{eq:30}--\eqref{eq:32}, and $B_i$ and $E_{ij}$ are independent, we have
\begin{align}
  \mathbb{E}\left[ S_B \right] &=  \mathbb{E}\left[S_{xxL} \sum_{i=1}^m \beta_i^2\right]
                                 \label{eq:33}
                             \\
                            &=  S_{xxL} \mathbb{E}\left[\sum_{i=1}^m \left(B_i -\bar{B}\right)^2 + \left( U_i -\bar{U}\right)^2\right]
                              \label{eq:34}
                              \\
   &= S_{xxL} (m-1)\sigma^2_B + (m-1)\sigma^2_E.
     \label{eq:35}
\end{align}
Therefore,
\begin{align}
  \mathbb{E}\left[V_B\right] =  \frac{\mathbb{E}\left[S_B\right]}{\phi_B} = S_{xxL} \sigma^2_B + (m-1) \sigma^2_E.
  \label{eq:36}
  \end{align}

\subsection{Derivation of $\mathbb{E}(V_R)$}
\label{sec:V_R}

Since $S_R = S_{xyT}^2 / S_{xxT}$, it follows that
\begin{align}
  S_R &=  \frac{1}{S_{xxT}} \left\{\sum_{i=1}^m \sum_{j=1}^n x_j \left[ (a_0 + A_i) + (b_0 + B_i) x_j + E_{ij}\right] \right\}^2
        \label{eq:37}\\
   &= \frac{1}{S_{xxT}} \left[ b_0  \sum_{i=1}^m \sum_{j=1}^n x_j^2 +  \sum_{i=1}^m B_{i} \sum_{j=1}^n x_j^2 +  \sum_{i=1}^m \sum_{j=1}^n x_j E_{ij} \right]^2
     \label{eq:38}\\
   &= S_{xxT} \left(b_0 +\frac{1}{m} \sum_{i=1}^m B_i + \frac{ \sum_{i=1}^m \sum_{j=1}^n x_j E_{ij} }{S_{xxT}} \right)^2.
     \label{eq:39}
\end{align}
From the independency of $B_i$ and $E_{ij}$, we have 
\begin{align}
  \mathbb{E} [S_R] = S_{xxT} \left(b_0^2 + \frac{1}{m^2} \sum_{i=1}^m \mathbb{E}\left[B_i^2\right] + \frac{1}{S_{xxT}^2} \sum_{i=1}^m \sum_{j=1}^n x_j^2 \mathbb{E}\left[E_{ij}^2\right]\right).
  \label{eq:40}
\end{align}
Here, since $B_i/\sigma_B \sim N(0,1)$, it follows that $B_i^2 \sim \sigma_B^2 \chi^2(1)$ and $\mathbb{E} [B_i^2] = \sigma_B^2$.
Also, since $\mathbb{E} [E_{ij}] = 0$, it follows that $\mathbb{E} [E_{ij}^2] = \mathbb{V} [E_{ij}] = \sigma^2_E$.

Thus, we have
\begin{align}
  \mathbb{E} [V_R] = \frac{\mathbb{E} [S_R]}{\phi_{R}} = S_{xxT} \left(b_0^2 + \frac{\sigma_B^2}{m} + \frac{\sigma_E^2}{S_{xxT}}\right) = b_0^2 S_{xxT} + \sigma_B^2 S_{xxT} + \sigma_E^2.
  \label{eq:41}
\end{align}

\subsection{Derivation of $\mathbb{E}(V_L)$}
\label{sec:V_L}

Since $S_L = S_A + S_B$, it follows that
\begin{align}
  \mathbb{E}[V_L] = \frac{\mathbb{E}[V_L]}{\phi_L} = \frac{n}{2}\sigma^2_A + \frac{S_{xxL}}{2} \sigma^2_B + \sigma^2_E.
  \label{eq:42}
\end{align}
From the definition of $\sigma^2_L$, it can be expressed using $\sigma^2_L$ as follows:
\begin{align}
 \mathbb{E}[V_L] = \frac{n}{2} \left(\sigma^2_A + \frac{S_{xxL}}{n}\sigma^2_B \right) + \sigma^2_E = \frac{n}{2} \sigma^2_L + \sigma^2_E.
  \label{eq:43}
\end{align}

\section{Model and precision measures in ISO 5725}
\label{sec:model-prec-meas-iso-}

The ISO 5725 series assumes that the measured values are quantitative data and its basic model is given by~\eqref{eq:44} (\cite{iso2025ISO572522025}):
\begin{equation}
	Y_{ij} = \mu  + L_i + E_{ij}, \quad i=1, \ldots, k, \ j=1, \ldots, n,
	\label{eq:44}
      \end{equation}
      where $Y_{ij}$ is the measured value of trial $j$ in laboratory $i$; $\mu$ is a general mean (expectation); $L_i$ is the laboratory component of variation (under repeatability conditions) in laboratory $i$, whose expectation is assumed to be $0$, and whose variance is the between-laboratory variance $\sigma_L^2$; and $E_{ij}$ is a residual error (under repeatability conditions), whose expectation is also assumed to be $0$, and whose variance is the within-laboratory variance $\sigma_{ri}^2$.
      We assume that $L_i$ and $E_{ij}$ are independent, and that the number of replicates is identical in all laboratories, denoted by $n$.
      Moreover, the within-laboratory variance $\sigma_{ri}^2$ is assumed to be identical across all laboratories and is denoted by the repeatability variance $\sigma_r^2$.
      The reproducibility variance $\sigma_R^2$ is defined as $\sigma_R^2 :=\sigma_L^2+\sigma_r^2$.
      It should be noted that the definition of reproducibility variance in the ISO 5725 series differs from that used in Gauge R\&R studies~\cite{burdick2005DesignAnalysisGauge}; the reproducibility variance in Gauge R\&R corresponds to the between-laboratory variance $\sigma_L^2$ in ISO 5725, but the definition in the ISO 5725 series is the first to be defined.
      
\end{document}